\documentclass{PoS}
\pdfoutput=1

\usepackage[utf8]{inputenc}
\usepackage{tikz}
\usepackage{amsmath}
\usepackage{amsfonts}
\usepackage{amssymb}
\usepackage{graphicx}
\usepackage{bbm}
\usepackage{braket}
\usepackage{xspace}
\usepackage[space]{grffile}

\newcommand{\comment}[1]{}

\renewcommand{\epsilon}{\varepsilon}

\renewcommand{\varphi}{\theta}

\newcommand{\dd}{D}
\newcommand{\one}{\mathbbm{1}}

\newcommand\qone{0+1d QCD\xspace}
\newcommand\qtwo{1+1d QCD\xspace}

\DeclareMathOperator\arsinh{arsinh}
\DeclareMathOperator\tr{tr}

\title{Complex Langevin in low-dimensional QCD: the good and the not-so-good}
\ShortTitle{Complex Langevin in low-dimensional QCD}

\author{\speaker{Jacques Bloch}\thanks{Supported by DFG}\\
       Institute for Theoretical Physics, University of Regensburg, Germany\\
       E-mail: \email{jacques.bloch@ur.de}}
\author{Johannes Mahr\\
       Institute for Theoretical Physics, University of Regensburg, Germany\\
       E-mail: \email{johannes.mahr@ur.de}}
\author{Sebastian Schmalzbauer\\
       Institute for Theoretical Physics, University of Regensburg, Germany\\
       E-mail: \email{sebastian.schmalzbauer@ur.de}}
       
\abstract{
We present our latest results on the application of the complex Langevin method to one- and two-dimensional QCD. Although the method is stable, it unfortunately converges to an incorrect result when applied as such. After applying additional gauge cooling steps, the results agree with the known analytical results in the one-dimensional case. However, in the two-dimensional case the disagreement subsists, even with gauge cooling, when the sign problem is sufficiently large.}

\FullConference{The 33rd International Symposium on Lattice Field Theory\\
		14 - 18 July 2015\\
		Kobe International Conference Center, Kobe, Japan}

\begin{document}

\section{Introduction}

Monte Carlo simulations of QCD at nonzero chemical potential are strongly hindered by the \textit{sign problem}, as the complex fermion determinant prohibits the use of importance sampling methods. Most known methods to circumvent the sign problem in QCD have a computational cost that grows exponentially with the volume. An alternative that has recently caught a lot of attention is the complex Langevin (CL) method \cite{Sexty:2013ica}. The CL stochastic differential equation uses the drift generated by the complex fermion action to evolve the complexified gauge configurations in the SL(3,$\mathbf{C}$) gauge group. After equilibration, and if a number of conditions are met \cite{Aarts:2011ax}, the time evolution of these configurations should reproduce the correct QCD results for gauge invariant observables.

Although the sign problem in QCD is particularly serious in four dimensions, it is already present in lower dimensions. In this talk we present a study on the viability of the CL method in 0+1d, where the sign problem is mild, and in 1+1d in the strong coupling limit where the sign problem is quite large in some regions of parameter space depending on the chemical potential, quark mass, temperature and spatial volume.

\section{Partition function and Dirac operator}

We consider the strong coupling partition function
\begin{align}
Z=\int\!\left[\prod_{x}\prod_\nu d U_{x,\nu}\right]\, \det D(\{U_{x,\nu}\})
\label{ZQCD}
\end{align}
with $d$-dimensional staggered Dirac operator
\begin{align}
 D_{xy}
 &= m \, \delta_{xy} + \frac12 \left[ e^{\mu} U_{x,0}\delta_{x+\hat{0},y}
   -e^{-\mu} U_{y,0}^{-1}\delta_{x-\hat{0},y}\right]
 +\frac12\sum_{i=1}^{d-1}\eta_i(x)\big[U_{x,i}\delta_{x+\hat i,y}-U_{y,i}^{-1}\delta_{x-\hat i,y}\big]
\label{eq:Dirac}
\end{align}
for a quark of mass $m$ at chemical potential $\mu$ and antiperiodic boundary conditions in the temporal direction. The staggered phases are $\eta_\nu=(-1)^{x_0+x_1+\ldots+x_{\nu-1}}$, $\hat\nu$ is a unit step in direction $\nu$, and we set the lattice spacing $a=1$. At zero $\mu$ the determinant of the Dirac operator is real and positive in SU(3), but at nonzero real $\mu$ the operator is no longer antihermitian, as $\dd(\mu)^\dagger=-\dd(-\mu)$, and its determinant becomes complex.

\section{Complex Langevin evolution}
\label{sec:0+1d-CLE}

We represent the gauge links using the Gell-Mann parameterization
\begin{align}
U = \exp\left[i\sum_a z_a \lambda_a \right] ,
\end{align}
with Gell-Mann matrices $\lambda_a$ and eight complex parameters $z_a$ for $U\in \text{SL}(3,\mathbf{C})$.

According to the CL equation, the discrete time evolution of $U_{x,\nu}$ in SL(3,$\mathbf{C}$) is given by the rotation \cite{Sexty:2013ica}
\begin{align}
U_{x,\nu}(t+1) = R_{x,\nu} (t) \: U_{x,\nu}(t) ,
\end{align}
where in the stochastic Euler discretization $R_{x,\nu} \in \text{SL}(3,\mathbf{C})$ is given by
\begin{align}
R_{x,\nu} = \exp\left[i\sum_a \lambda_a (\epsilon K_{a,x,\nu} + \sqrt{\epsilon}\,\eta_{a,x,\nu})\right] ,
\end{align}
with drift term
\begin{align}
K_{a,x,\nu} &= - D_{a,x,\nu}(S)
= - \partial_\alpha S(U_{x,\nu}\to e^{i\alpha\lambda_a} U_{x,\nu})|_{\alpha=0} ,
\end{align}
real Gaussian noise $\eta_{a,x,\nu}$ with variance 2, fermion action $S=-\log \det D$ and discrete Langevin time step $\epsilon$.

\section{Gauge cooling}
\label{sec:gc}

Previous studies using the CL method have shown that incorrect results are obtained when the simulation wanders off too far in the imaginary direction.  In gauge theories it was suggested to counter this problem  using \textit{gauge cooling}, where the SL(3,$\mathbf{C}$) gauge invariance of the theory is used to keep the trajectories as closely as possible to the SU(3) group \cite{Seiler:2012wz}. 

A general gauge transformation of the link $U_{x,\nu}$ is given by
\begin{align}
U_{x,\nu} \to G_x \, U_{x,\nu} \, G_{x+\hat{\nu}}^{-1} 
\label{gc}
\end{align}
with $G_x \in \text{SL(3,$\mathbf{C}$})$. Gauge cooling corresponds to the minimization of the unitarity norm
\begin{align}
 ||\mathcal{U}|| = \sum_{x,\nu} \tr \left[ U_{x,\nu}^{\dagger}U_{x,\nu} + \left(U_{x,\nu}^{\dagger}U_{x,\nu}\right)^{-1} \!\!-2\right] 
\label{unorm}
\end{align}
over all $G_x$, which is usually done via steepest descent.

Clearly, observables are \textit{invariant} under gauge transformations and so is the drift term in the CL equation. However, as the noise distribution in the CL equation is \textit{not invariant} under SL(3,$\mathbf{C}$) gauge transformations, the gauge cooling and Langevin steps do not commute, which leads to different trajectories in configuration space when cooling is introduced. Although this is exactly the aim of the cooling procedure, 
it is still an open question whether or under what conditions this procedure leads to the correct QCD expectation values (see also \cite{Nagata:2015uga} for recent developments).

\section{QCD in 0+1 dimensions}

We first consider \qone where the determinant of the Dirac operator \eqref{eq:Dirac} can be reduced to the determinant of a $3\times 3$ matrix \cite{Bilic:1988rw}
\begin{align}
  \det D \propto \det \left[ e^{\mu/T} P+e^{-\mu/T} P^{-1} + 2\cosh\left(\mu_c/T\right)\,\one_3\right]
\end{align}
with Polyakov line $P= \prod_t U(t)$ and effective mass $\mu_c = \arsinh(m)$. The partition function is then a one-link integral of $\det D$ over $P$ without gauge action. As analytic results  \cite{Bilic:1988rw, Ravagli:2007rw}, as well as numerical solutions using subsets \cite{Bloch:2013ara}, are available in this case, the correctness of the numerical results obtained with the CL method can be verified.

Note that some modified models for \qone were already solved using the CL method, including a one-link formulation with mock-gauge action \cite{Aarts:2008rr} and a U($N_c$) theory in the spectral representation \cite{Aarts:2010gr}.

In \qone gauge transformations \eqref{gc} simplify to
\begin{align}
P \to  G P G^{-1} ,
\end{align}
only depending on a single $G \in \text{SL(3,$\mathbf{C}$)}$. It is easy to show that in this case maximal cooling, i.e.\ minimizing \eqref{unorm}, is achieved by the similarity transformation diagonalizing $P$. We found that cooling typically reduces the unitarity norm by about two orders of magnitude.

\begin{figure}[t]
\begin{center}
\parbox{0.33\textwidth}{
\includegraphics[width=0.32\textwidth]{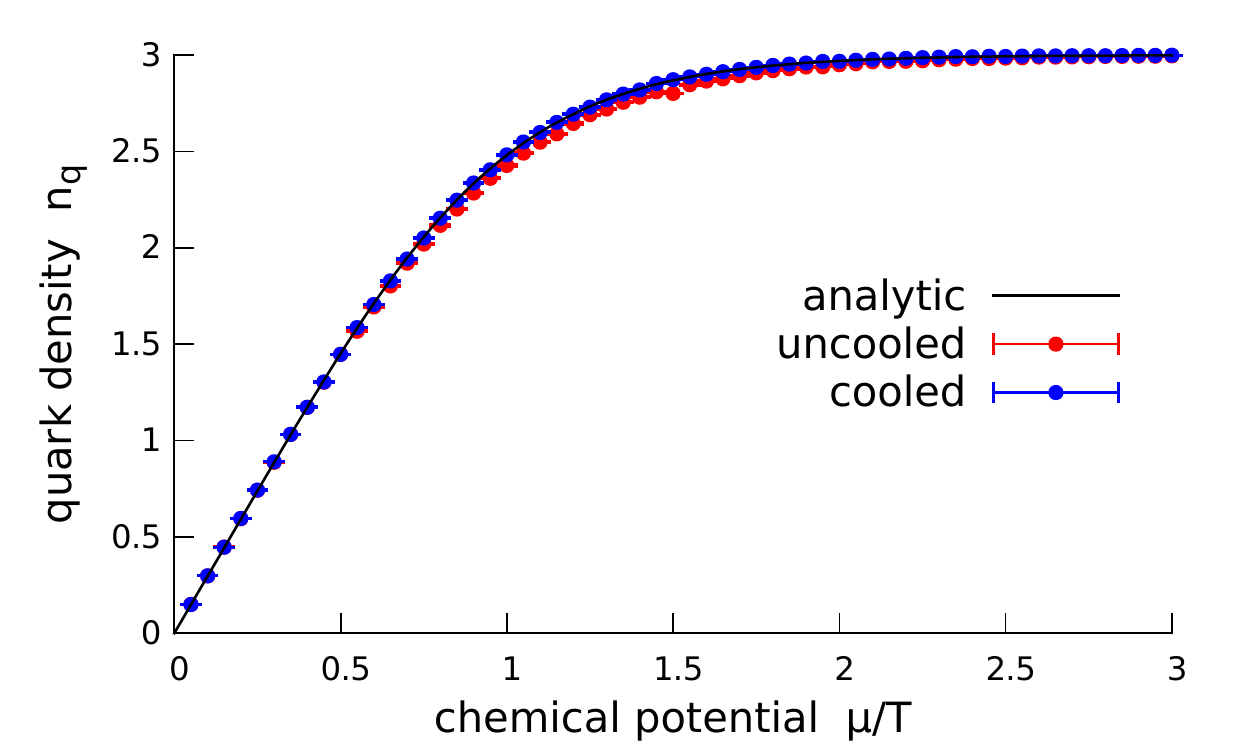}\\
\hspace*{1mm}\includegraphics[width=0.32\textwidth]{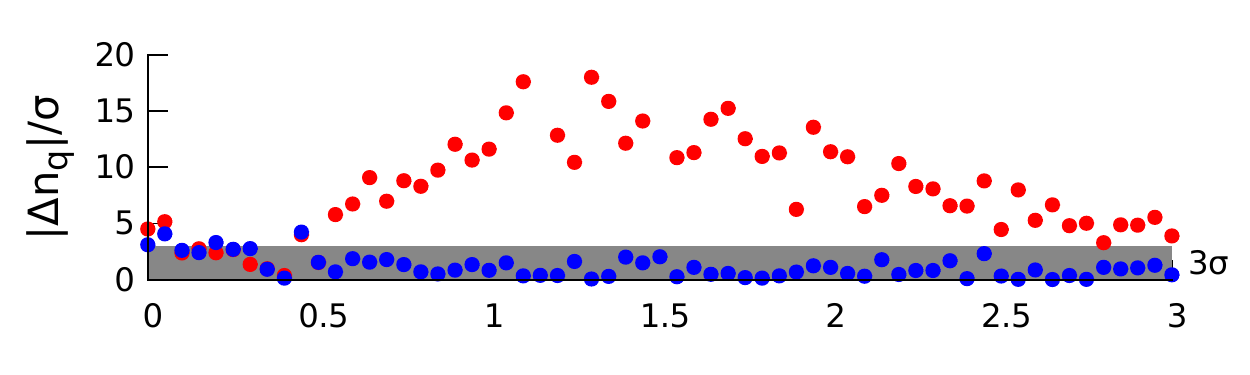}
}
\hspace{5mm}
\parbox{0.33\textwidth}{
\includegraphics[width=0.32\textwidth]{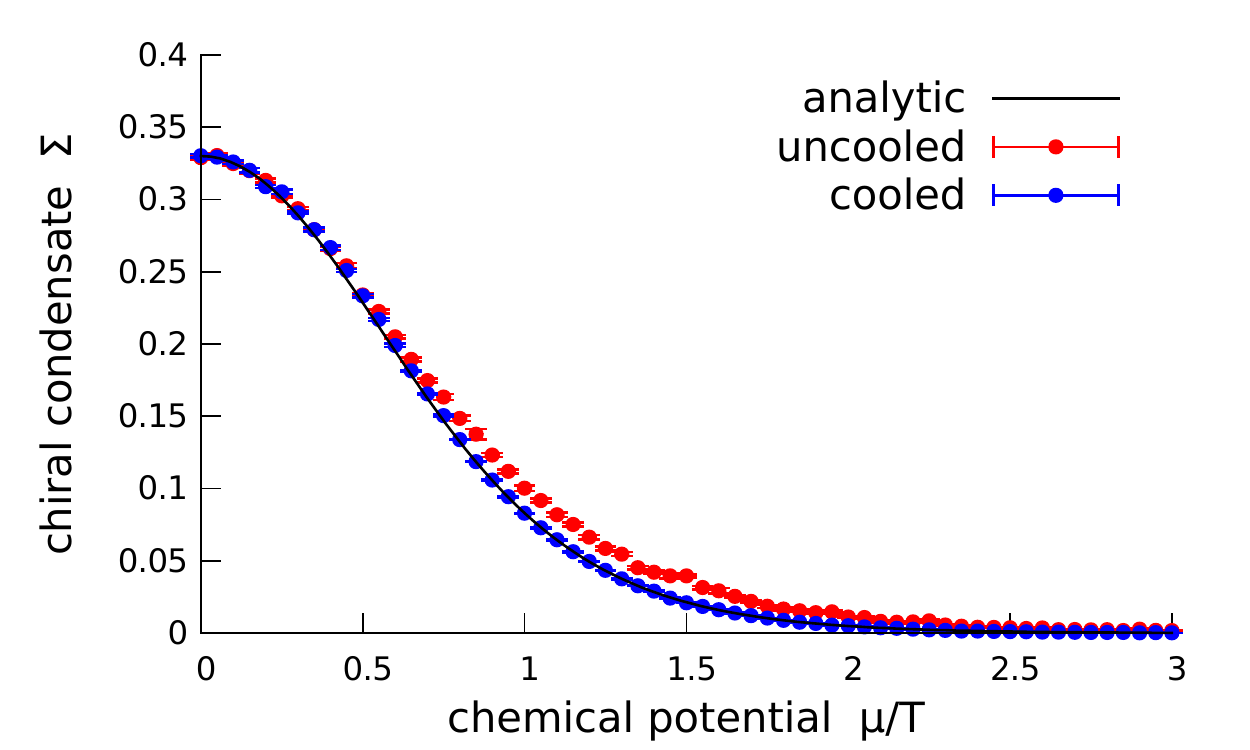}\\
\hspace*{2mm}\includegraphics[width=0.32\textwidth]{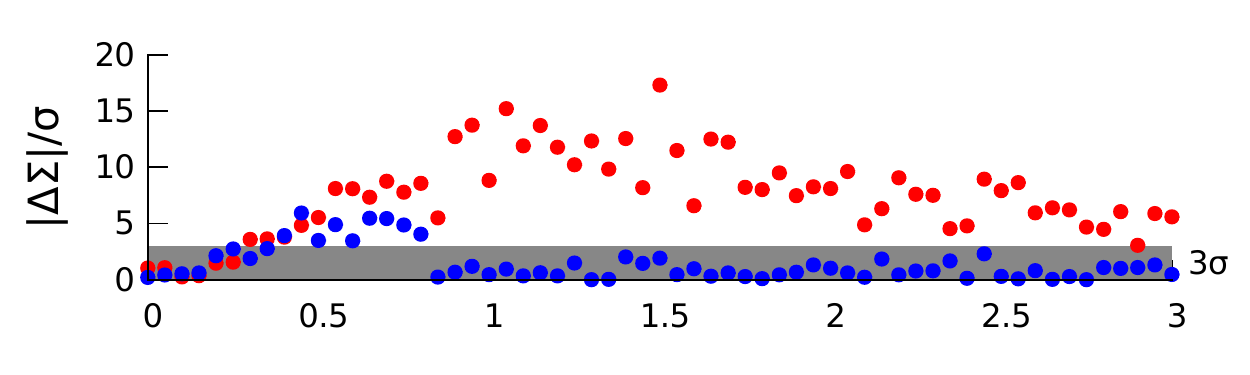}
}
\end{center}
\vspace{-6mm}
\caption{Quark density and chiral condensate as a function of $\mu/T$ for $m=0.1$: data (top row) and statistical significance of the deviation between numerical and analytical results (bottom row).\vspace{-2mm}}
\label{fig3}
\end{figure}
In Fig.\ \ref{fig3} we show the results for the quark density and chiral condensate as a function of $\mu/T$ for $m=0.1$. Below the data we show the statistical significance of the deviation between the numerical result $y$ and the analytical result $y_\text{th}$, i.e.\ $|y-y_\text{th}|/\sigma_y$. For the uncooled results  the deviation is far too large to be attributed to statistical fluctuations and we conclude that the CL method introduces a systematic error. After cooling, however, the CL results are in agreement with the theoretical predictions within the statistical accuracy (except for $\mu \approx 0.7$ where the deviation is still somewhat too large for the chiral condensate). Gauge cooling seems absolutely necessary to get the correct result, even in this one-dimensional gauge theory.

\begin{figure}[t]
\centerline{
\includegraphics[height=0.25\linewidth]{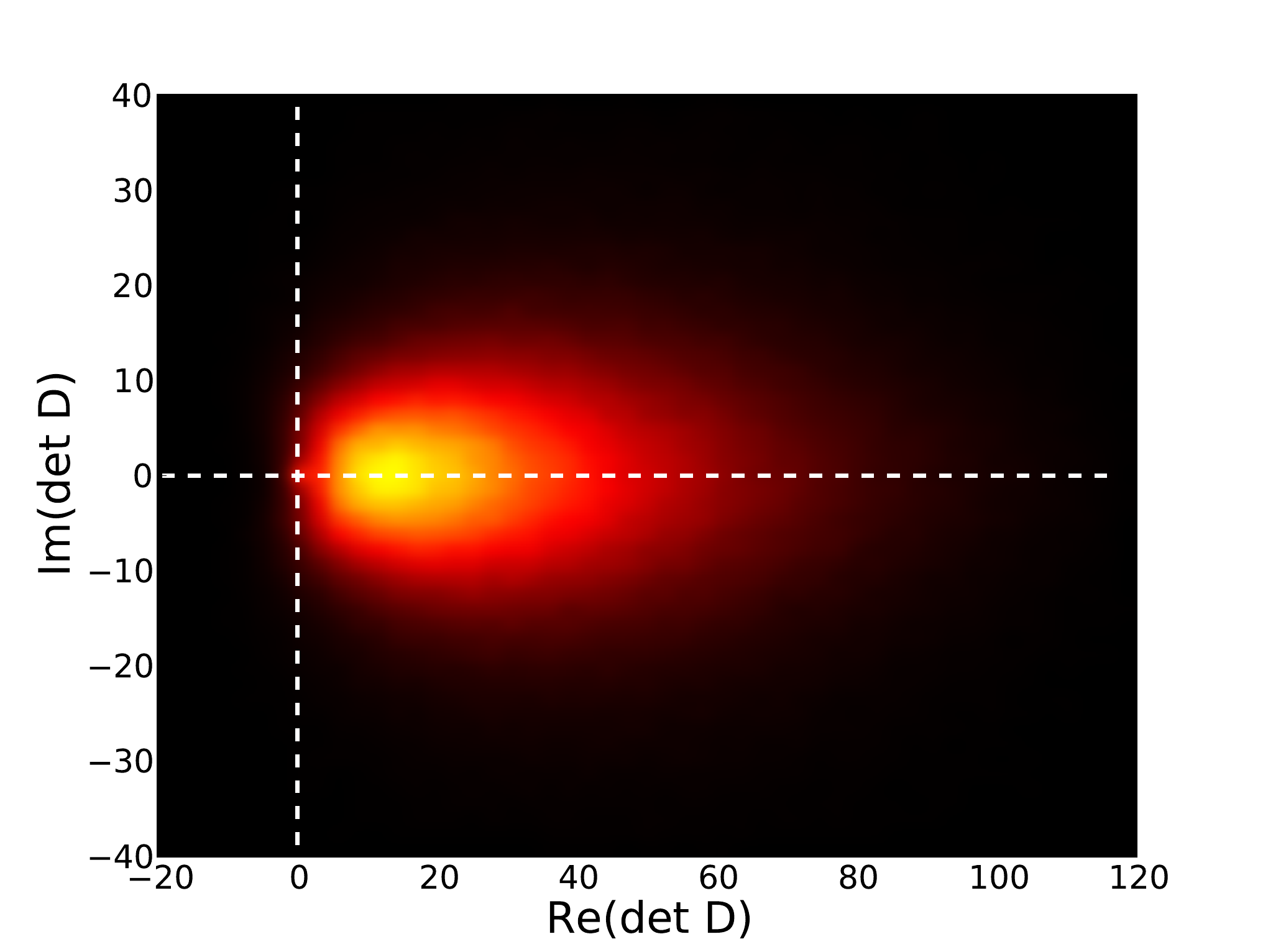}
\hspace{5mm}
\includegraphics[height=0.25\linewidth]{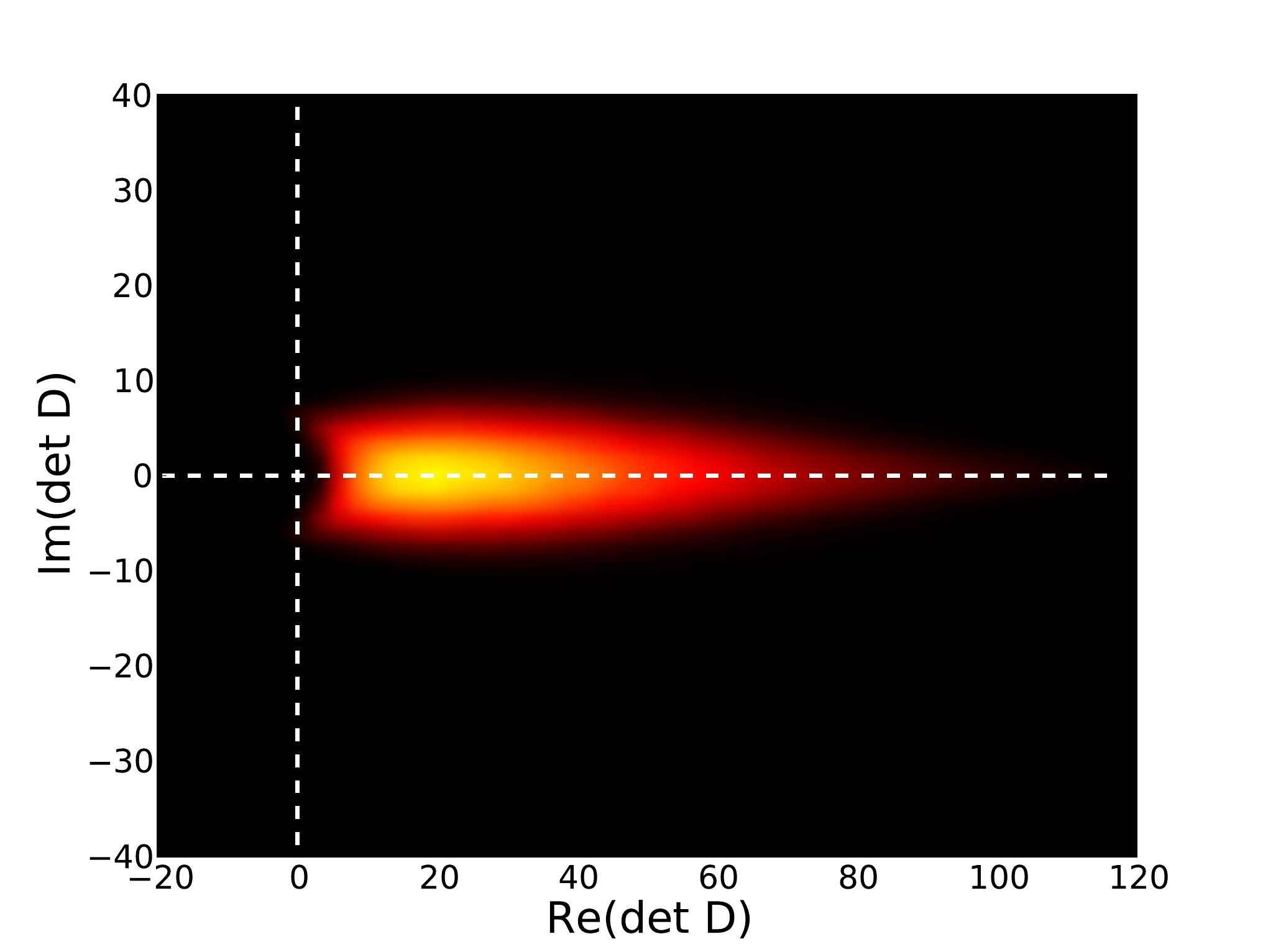}
}
\vspace{-2mm}
\caption{Density of the value of the determinant in the complex plane for $\mu/T=1$ and $m=0.1$ for the uncooled (left) and cooled (right) cases.\vspace{-2mm}}
\label{fig4}
\end{figure}
To illustrate the effect of gauge cooling on the SL(3, $\mathbf{C}$) trajectories we show how the density of the determinant in the complex plane is affected by cooling in Fig.\ \ref{fig4}. The effect is quite dramatic, as the origin, which is inside the distribution without cooling, is clearly avoided when cooling is applied. Avoiding the singular drift at the origin could be a necessary condition for the complex Langevin to yield the correct result \cite{Aarts:2011ax,Nishimura:2015pba}.

\begin{figure}[t]
\centerline{
\includegraphics[width=0.32\textwidth]{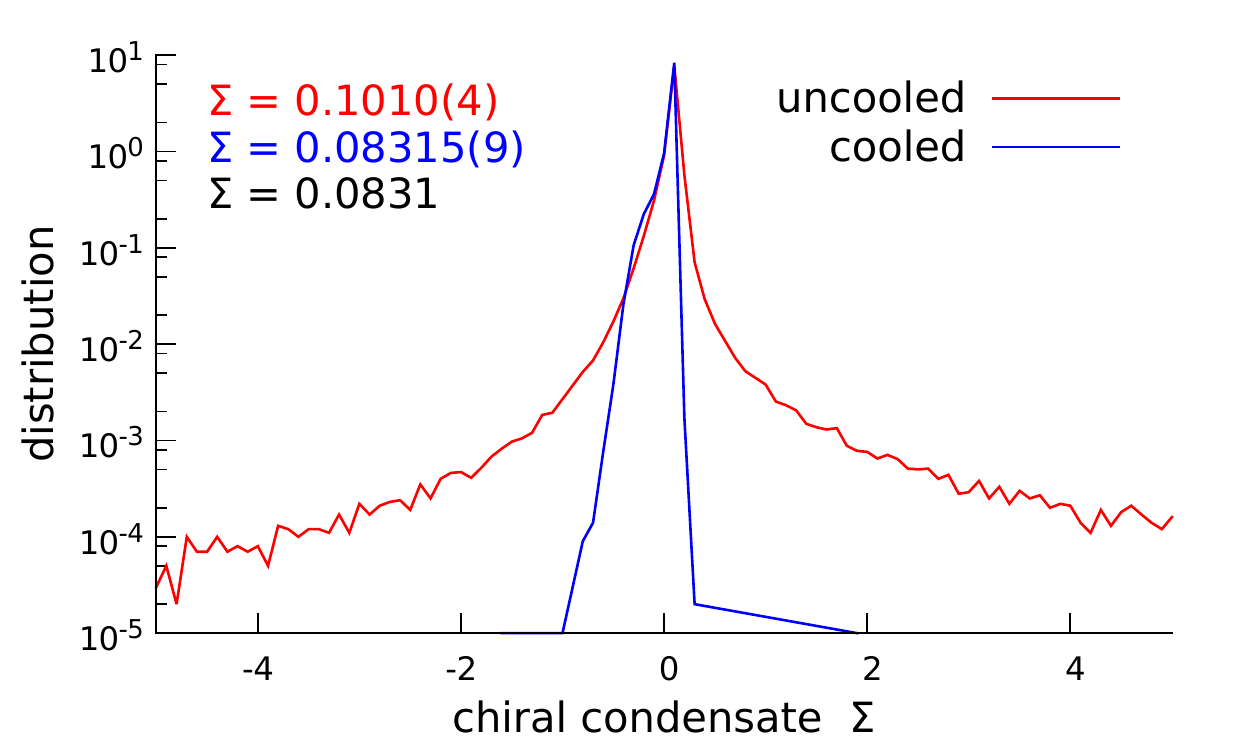}
\hspace{5mm}
\includegraphics[width=0.32\textwidth]{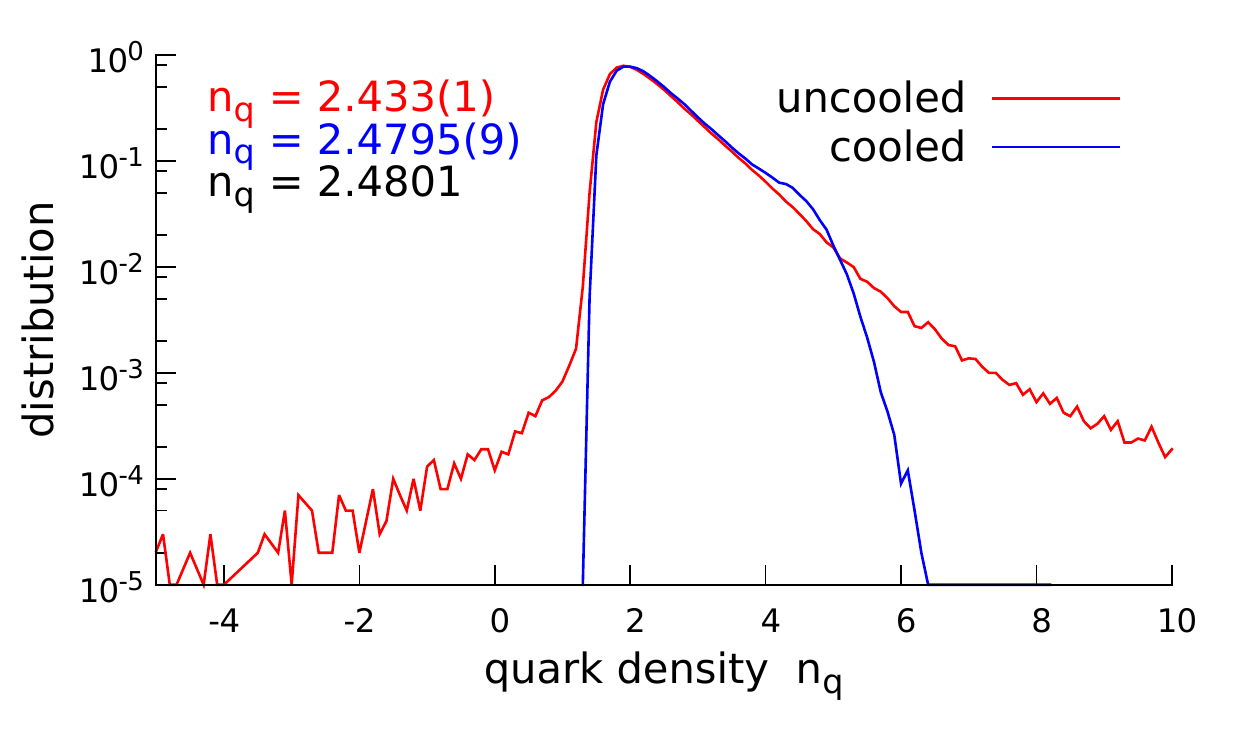}
}
\vspace{-2mm}
\caption{
Distribution of the real part of the chiral condensate (left) and quark number density (right) at $\mu/T=1$ and $m=0.1$ in the uncooled (red) and cooled (blue) cases.\vspace{-2mm}}
\label{fig5}
\end{figure}

Another known signal for problems in the CL method is the existence of skirts in the distribution of the (real part of the) observables \cite{Aarts:2013uxa}. This is illustrated in Fig.\ \ref{fig5}. Without cooling the observables have very wide skirts, hinting at a polynomial decay, while after cooling very sharp exponential fall offs are observed.

Note that in the 0+1d case we can also parameterize the Polyakov line in its diagonal representation with two complex parameters. The numerical results obtained with the CL method in this representation agree with the analytical predictions. This is consistent with the above results as gauge cooling also brings the Polyakov line to its diagonal form in the Gell-Mann parameterization.

\section{QCD in 1+1 dimensions}

A more stringent test of the CL method is provided by \qtwo where the sign problem is more severe. The staggered Dirac operator is given in \eqref{eq:Dirac} and in this work we restrict our simulations to the strong coupling case. The complex Langevin equations are given in Sec.\ \ref{sec:0+1d-CLE} and the gauge cooling procedure in Sec.\ \ref{sec:gc}.

All results shown here are preliminary and were obtained on a $4\times 4$ lattice. We are currently performing further evaluation runs for lattices of size $N_s\times N_t=4\times\{2,6,8,10\}$, $6\times\{2,4,6,8\}$ and $8\times\{2,4,6,8\}$.

\begin{figure}[t]
\centerline{
\includegraphics[width=0.32\textwidth]{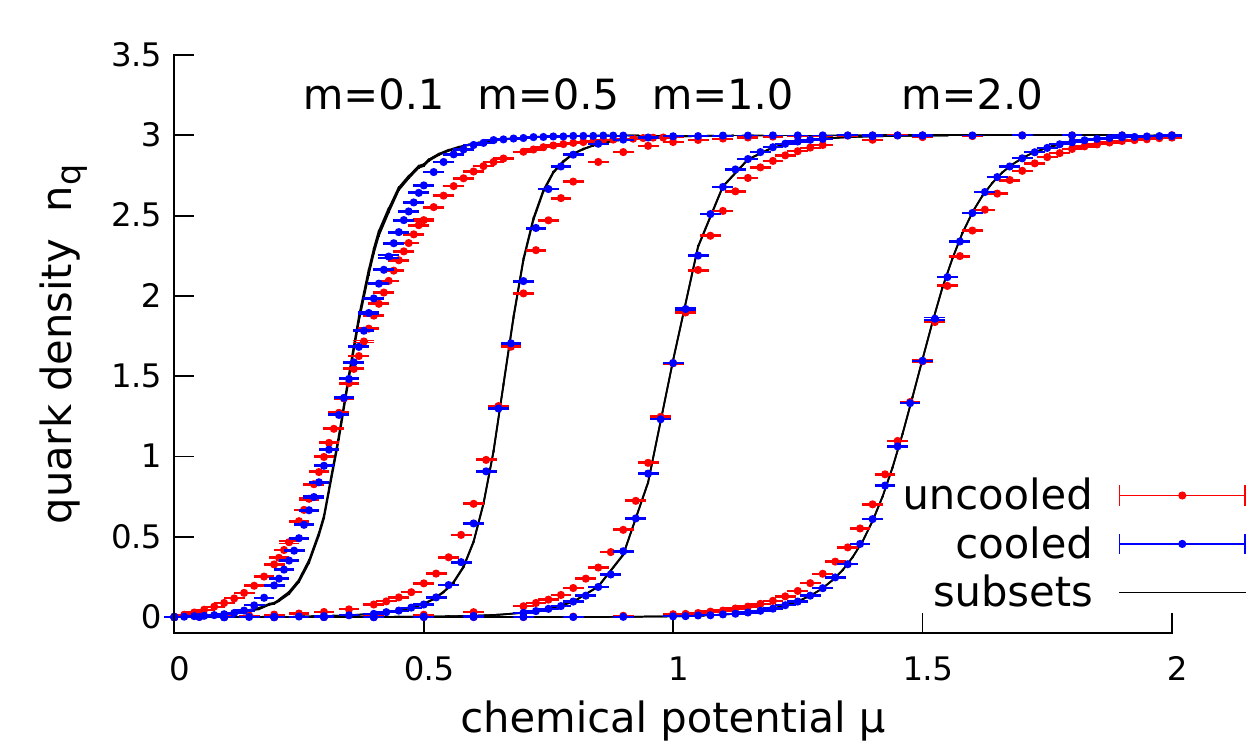}
\hspace{8mm}
\includegraphics[width=0.32\textwidth]{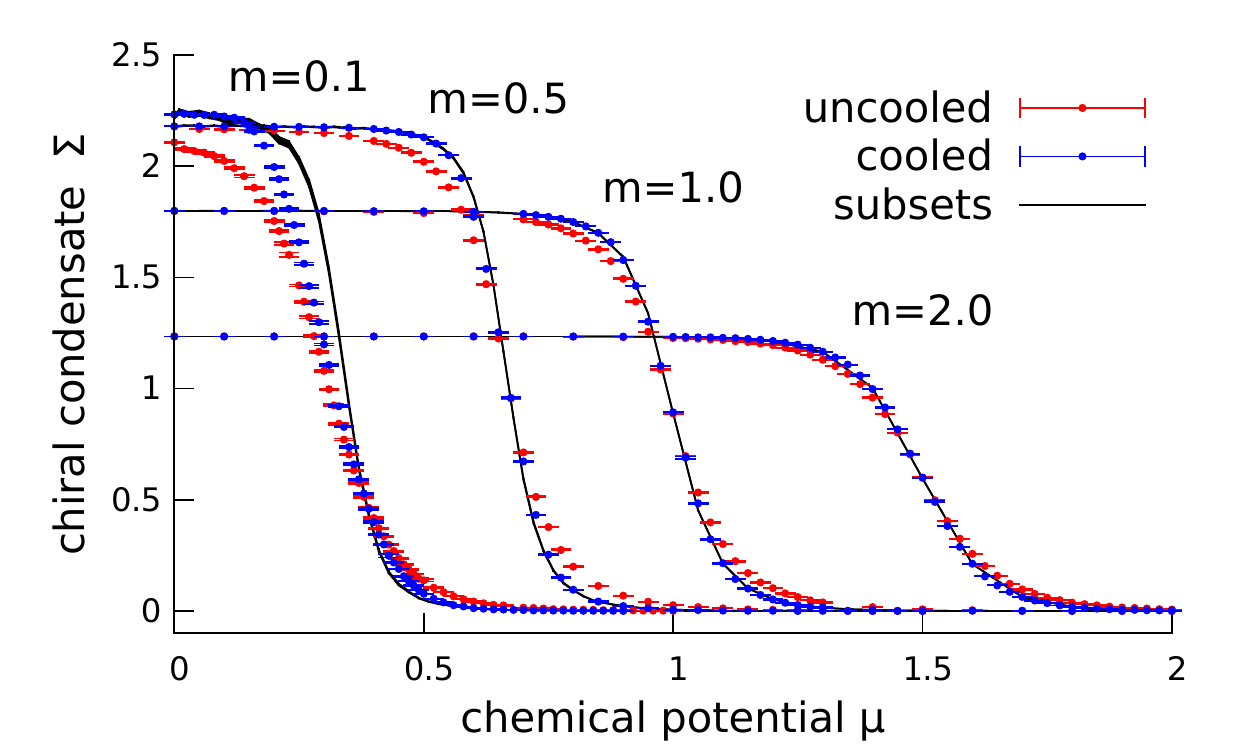}
}
\vspace{-2mm}
\caption{Quark density (left) and chiral condensate (right) as a function of the chemical potential for $m=0.1,0.5,1.0,2.0$. We compare uncooled (red) and cooled (blue) CL results with subset results (line).\vspace{-2mm}}
\label{fig8}
\end{figure}

To validate the CL method we compare our CL measurements with results obtained using the subset method \cite{Bloch:2013qva, Bloch:2015iha}. As can be seen in Fig.\ \ref{fig8}, the results of the bare or uncooled CL are not consistent with the subset data, in all cases considered. After cooling the situation is much improved and the large mass CL simulations agree very well with the subset results over the complete $\mu$-range. However, for the smallest mass value ($m=0.1$), even the cooled CL does not produce correct results over a large range of $\mu$ values. Furthermore, a close inspection of the $m=0.5$ results also shows a significant deviation. Clearly, the CL does not work for light quarks even after applying full gauge cooling. 

\begin{figure}[t]
\begin{center}
	\includegraphics[height=0.25\linewidth]{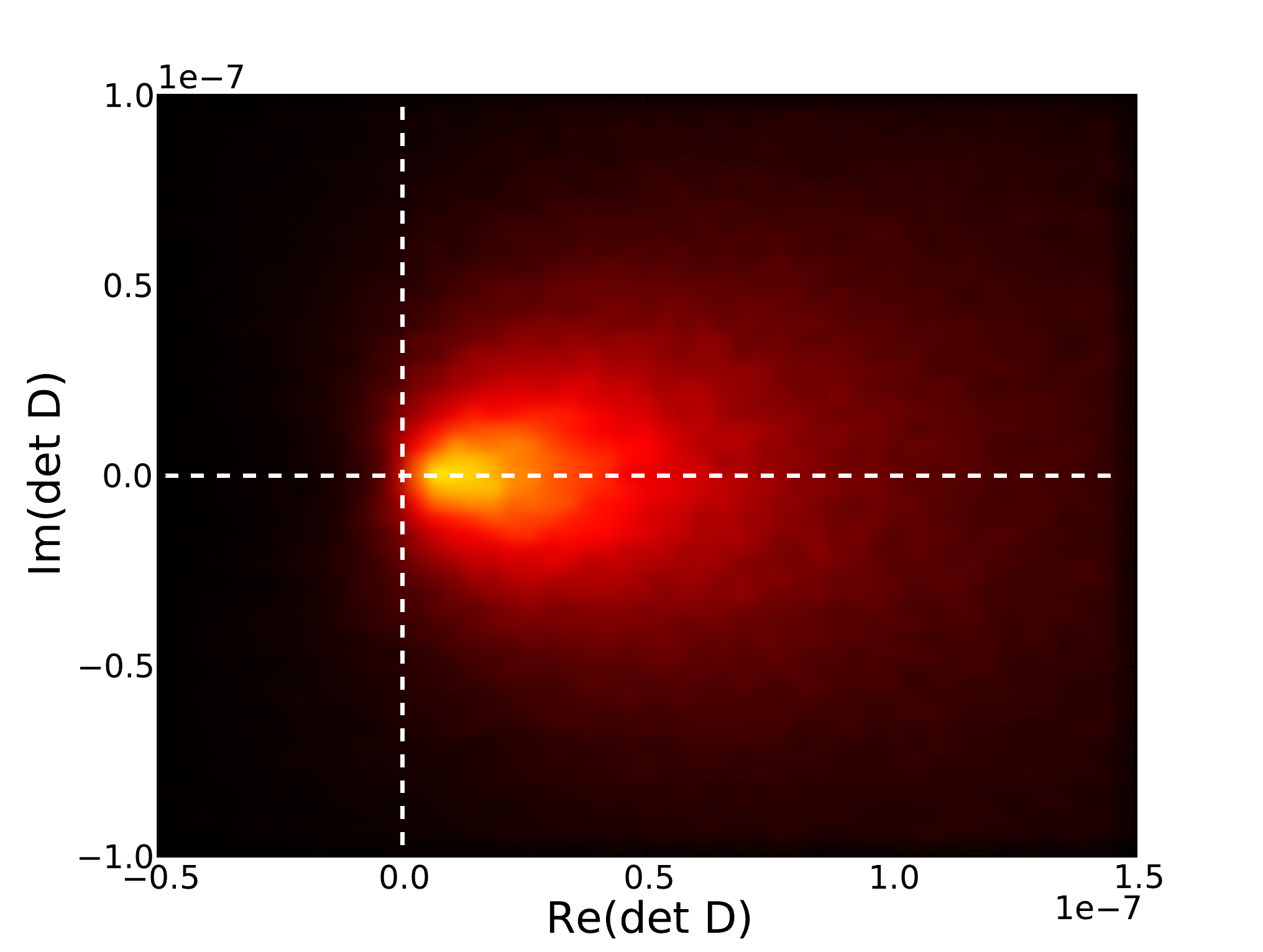}
	\hspace{5mm}
	\includegraphics[height=0.25\linewidth]{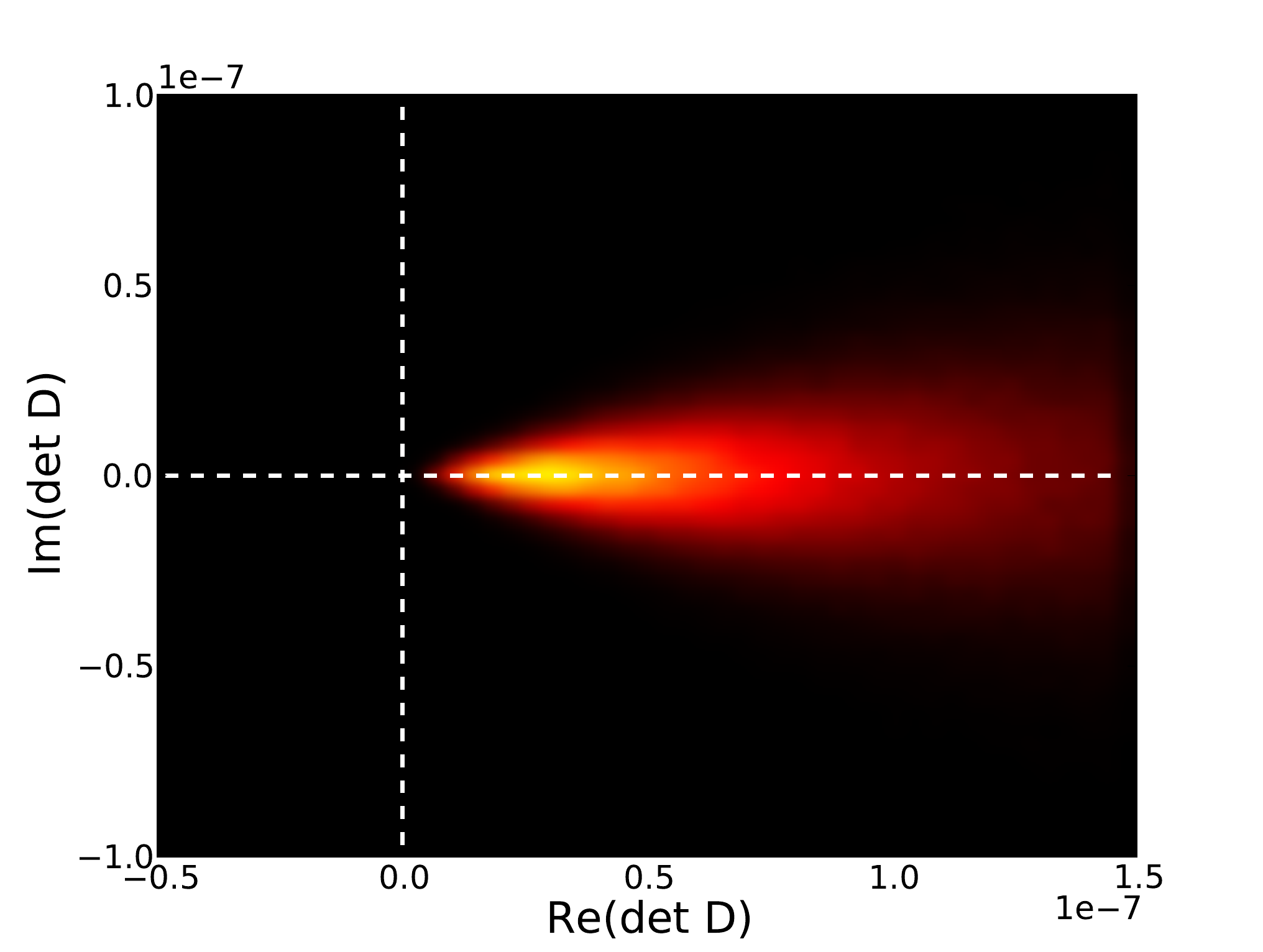}
\\[-0.1cm]
	\includegraphics[height=0.25\linewidth]{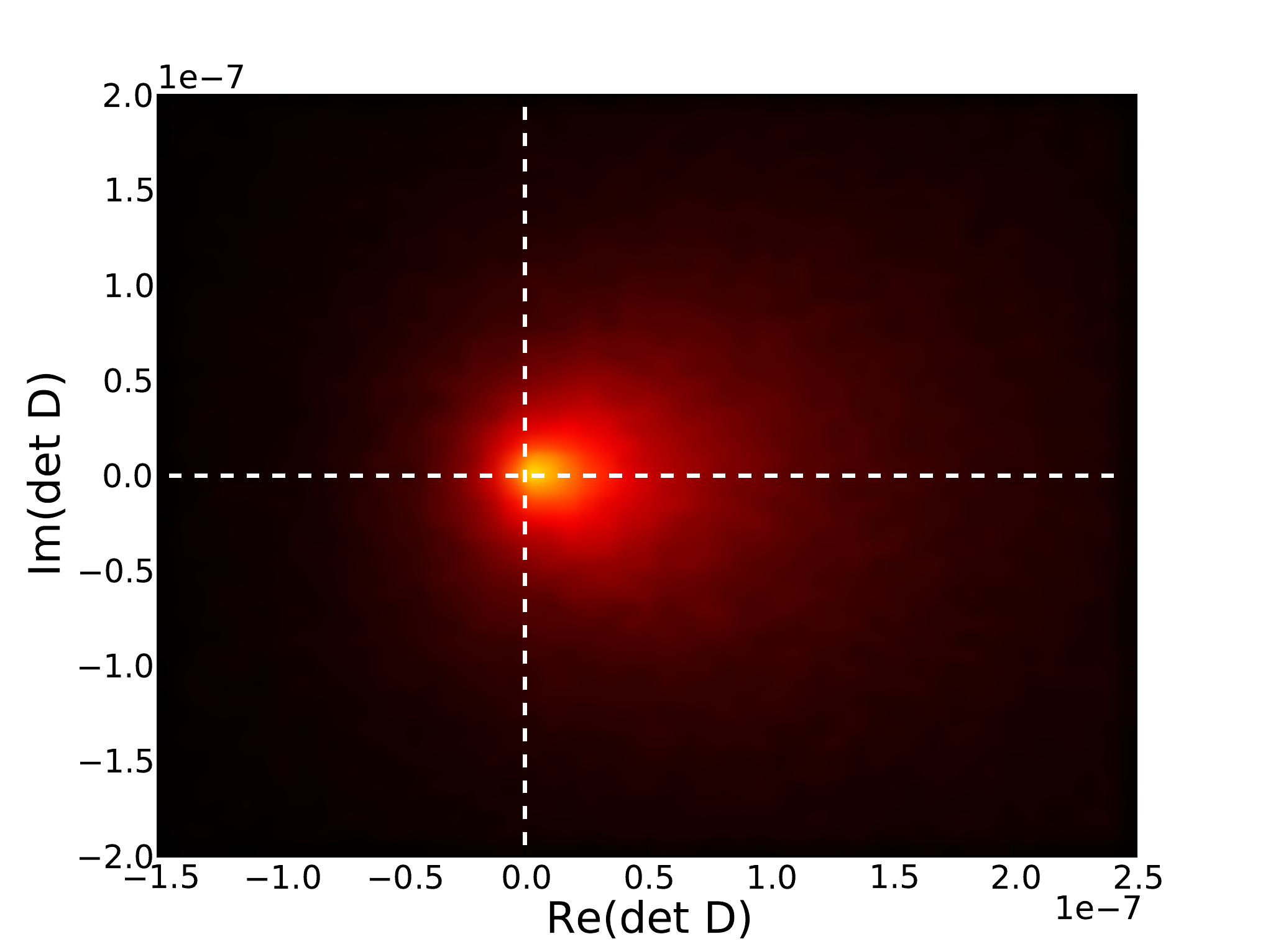}
	\hspace{5mm}
	\includegraphics[height=0.25\linewidth]{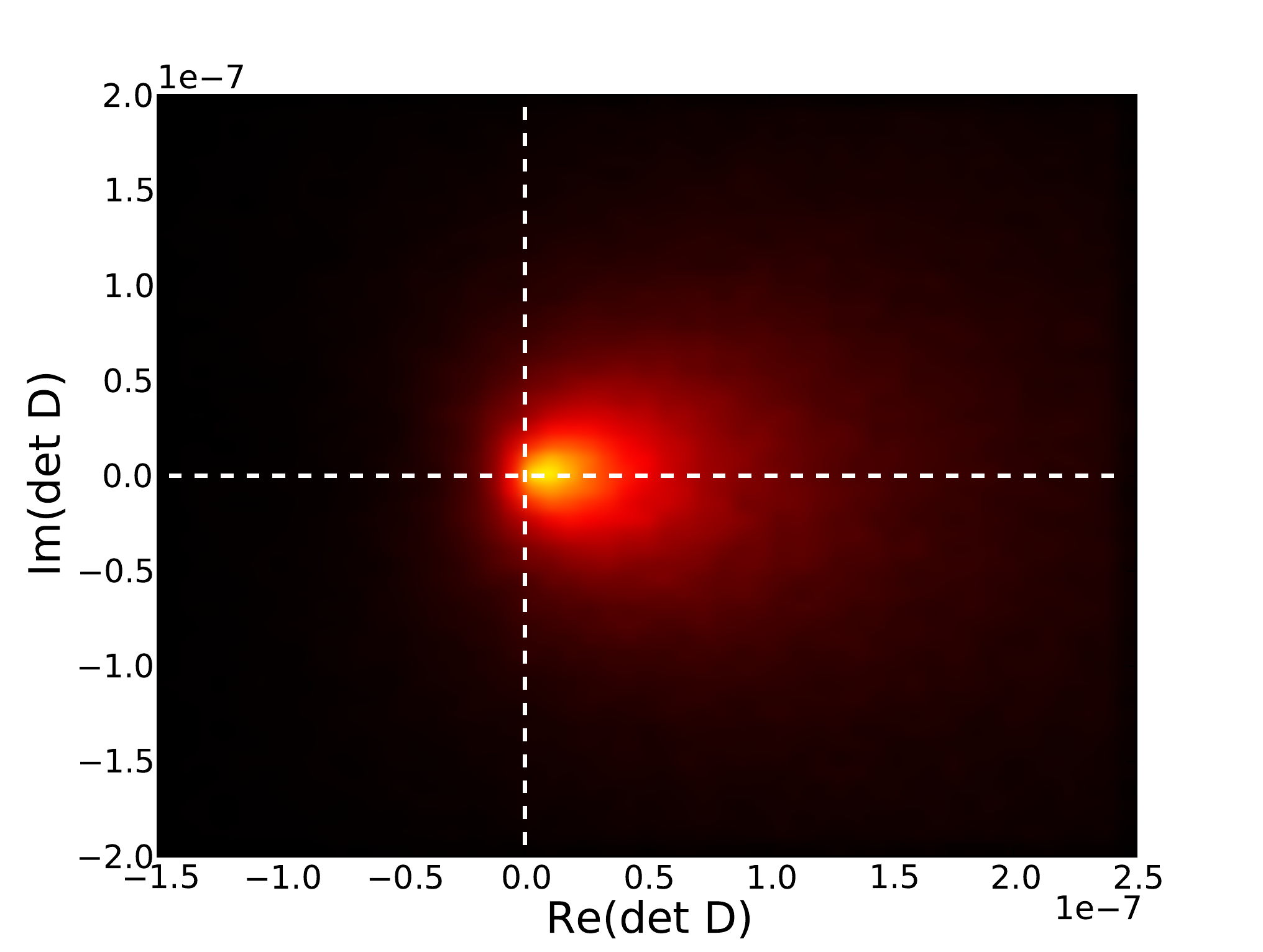}
\end{center}
\vspace{-6mm}
\caption{Density distribution of the determinant value for $\mu=0.07$ (top) and $\mu=0.25$ (bottom) in the uncooled (left) and cooled (right) case.\vspace{-2mm}}
\label{fig10}
\end{figure}

To investigate why gauge cooling does not work for small masses, we look at its effect on the density distribution of the determinant for $m=0.1$. In Fig.\ \ref{fig10} we see that for $\mu=0.07$, where the CL results seem correct, cooling significantly changes the distribution: it squeezes the density along the real axis while also pushing it away from the origin. For $\mu=0.25$, however, cooling has very little effect: the fireball is somewhat shifted to the right but its shape remains approximately unchanged and the origin is still inside the distribution. The CL results are thus incorrect when cooling is unable to change the distribution substantially, such that it still contains the origin and remains broad in the imaginary direction.

\begin{figure}[t]
\centerline{
\includegraphics[width = .32\textwidth]{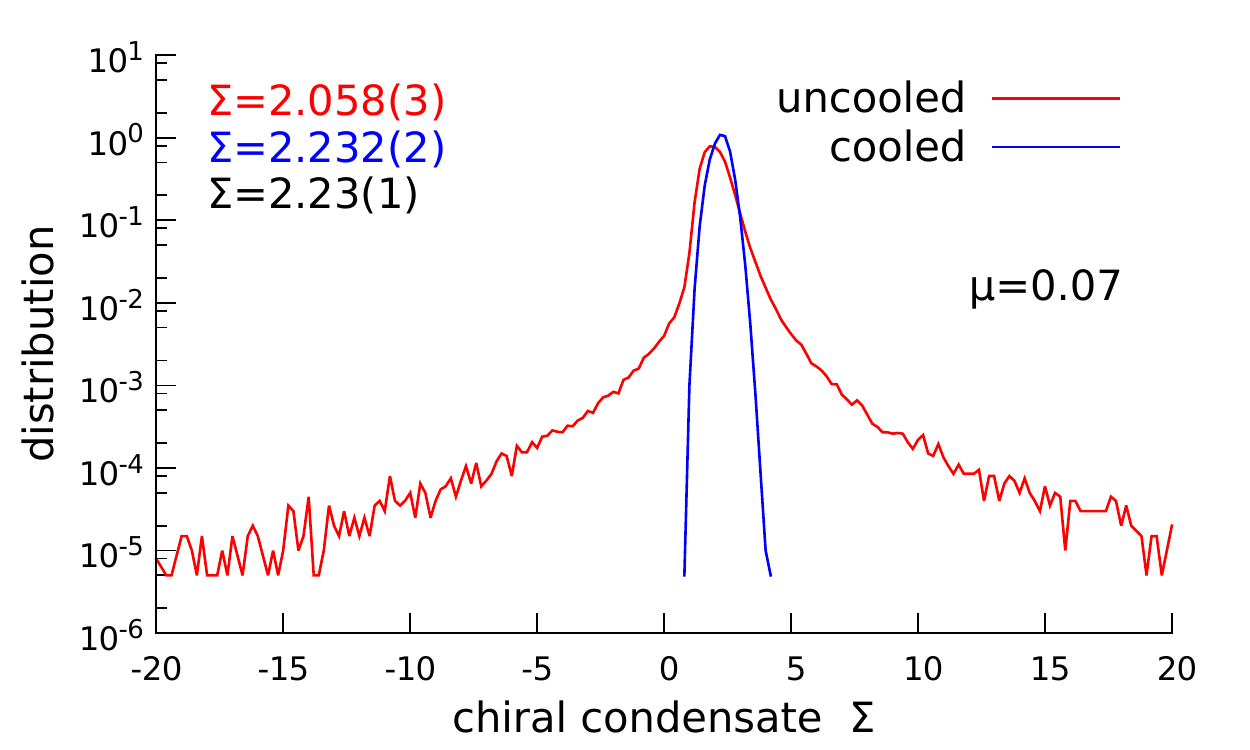}
\includegraphics[width = .32\textwidth]{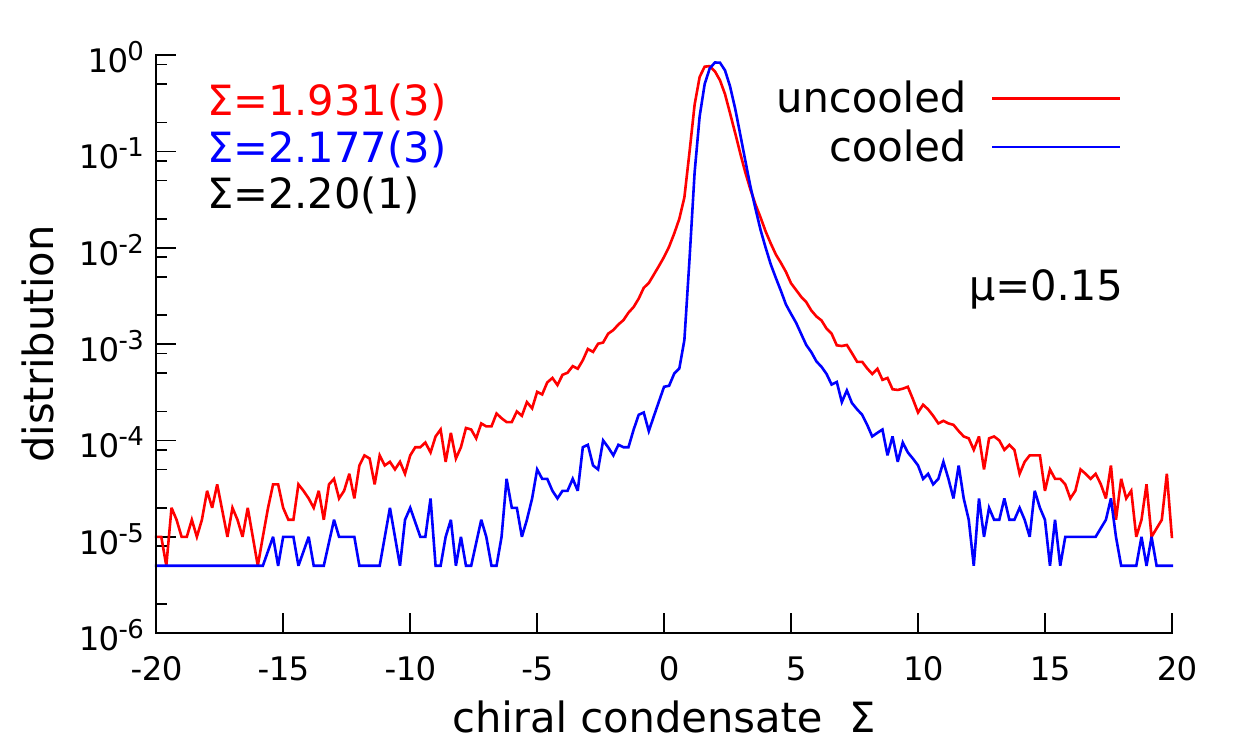}
\includegraphics[width = .32\textwidth]{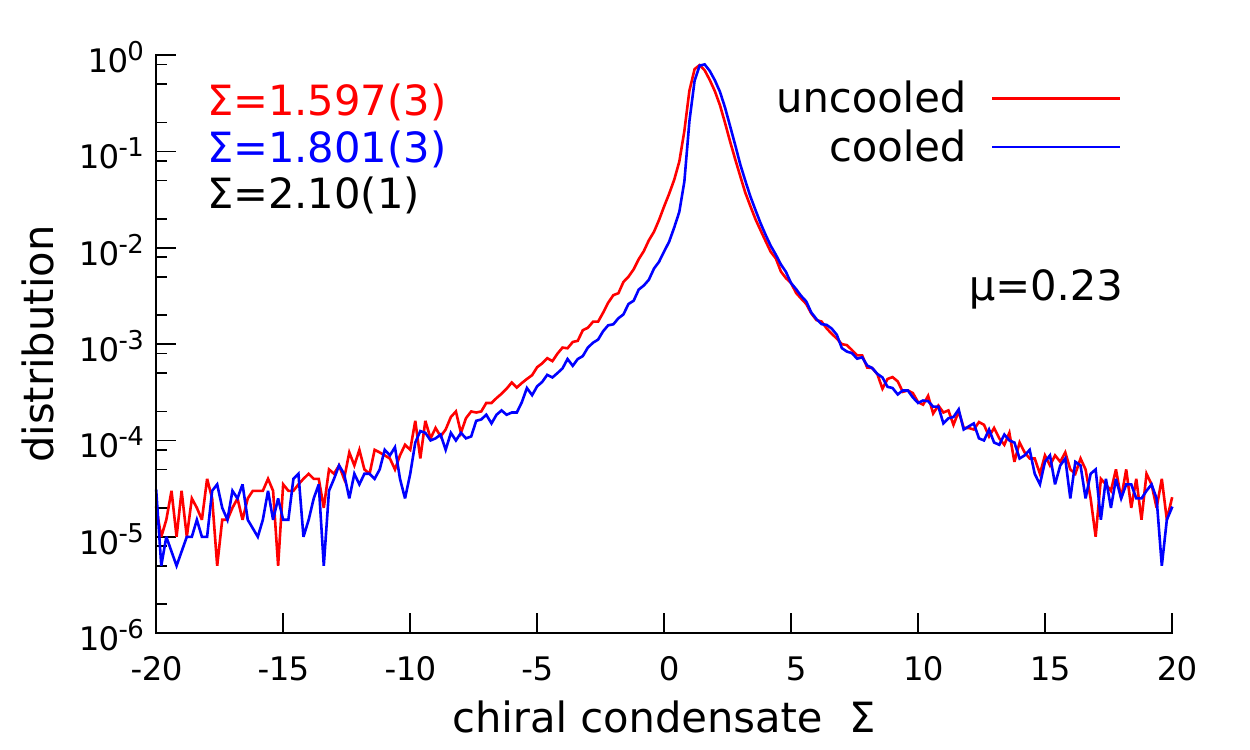}
}
\vspace{-1mm}
\caption{Distribution of the real part of the chiral condensate for $\mu=0.07,0.15,0.23$ for $m=0.1$ without (red) and with (blue) cooling.\vspace{-2mm}}
\label{fig11}
\end{figure}

We also looked at the effect of cooling on the distribution of the observables. This is illustrated in Fig.\ \ref{fig11}, which shows the distribution of the real part of the chiral condensate for increasing chemical potential, without and with cooling for $m=0.1$. The uncooled distribution always displays skirts, with a decay that is fairly independent of $\mu$. With gauge cooling we observe that the skirt vanishes for small chemical potentials, but as it is increased the skirts gradually reappear, signaling that the results of the CL method gradually become untrustworthy for light quarks, even in the presence of cooling, as we move into the region that has a substantial sign problem.

\section{Conclusions}

In this work we have shown that in \qone the results obtained with the complex Langevin method deviate significantly from the analytical predictions when no gauge cooling is applied. After introducing gauge cooling the correct results are recovered. 

In \qtwo at strong coupling the uncooled CL method yields wrong results for any mass and chemical potential. When applying gauge cooling the results are rectified for heavy quarks, but for light quarks the results remain incorrect for a significant range of the chemical potential.

Gauge cooling seems absolutely necessary, although not sufficient in some cases. As the quarks get lighter and the sign problem larger, gauge cooling no longer works properly. The results were validated by comparing with subset measurements, but signals for wrong convergence are also available within the CL method itself, such as skirts in observable distributions, and distributions of the determinant that contain the origin and are broad in the imaginary direction, even after cooling. 

From our study we conclude that much remains to be understood about the complex Langevin method and its applicability to QCD.  Several suggestions presented in the literature such as changes of variables \cite{Mollgaard:2014mga} or using different cooling norms or gauge fixing conditions \cite{KeitaroNagata2015} should be investigated. 
We are currently validating the CL method for light quarks on larger lattices and in the presence of a gauge action, as it is believed that the CL method performs better in the weak coupling regime. Several unanswered questions remain: does the CL method work when the sign problem is large and how well can we trust its results considering that its degradation seems to happen gradually and not in an on-off way?

\sloppypar
\bibliography{biblio}

\providecommand{\href}[2]{#2}\begingroup\raggedright\begin{thebibliography}{10}

\bibitem{Sexty:2013ica}
D.~Sexty,\href{http://dx.doi.org/10.1016/j.physletb.2014.01.019}{ {\em Phys.
  Lett.} {\bf B729} (2014) 108} [\href{http://arxiv.org/abs/1307.7748}{{\tt
  arXiv:1307.7748}}].

\bibitem{Aarts:2011ax}
G.~Aarts, F.~A. James, E.~Seiler, and I.-O.
  Stamatescu,\href{http://dx.doi.org/10.1140/epjc/s10052-011-1756-5}{ {\em
  Eur.Phys.J.} {\bf C71} (2011) 1756}
  [\href{http://arxiv.org/abs/1101.3270}{{\tt arXiv:1101.3270}}].

\bibitem{Seiler:2012wz}
E.~Seiler, D.~Sexty, and I.-O.
  Stamatescu,\href{http://dx.doi.org/10.1016/j.physletb.2013.04.062}{ {\em
  Phys.Lett.} {\bf B723} (2013) 213}
  [\href{http://arxiv.org/abs/1211.3709}{{\tt arXiv:1211.3709}}].

\bibitem{Nagata:2015uga}
K.~Nagata, J.~Nishimura, and S.~Shimasaki,
  \href{http://arxiv.org/abs/1508.02377}{{\tt arXiv:1508.02377}}.

\bibitem{Bilic:1988rw}
N.~Bilic and
  K.~Demeterfi,\href{http://dx.doi.org/10.1016/0370-2693(88)91240-3}{ {\em
  Phys.Lett.} {\bf B212} (1988) 83}.

\bibitem{Ravagli:2007rw}
L.~Ravagli and
  J.~Verbaarschot,\href{http://dx.doi.org/10.1103/PhysRevD.76.054506}{ {\em
  Phys.Rev.} {\bf D76} (2007) 054506}
  [\href{http://arxiv.org/abs/0704.1111}{{\tt arXiv:0704.1111}}].

\bibitem{Bloch:2013ara}
J.~Bloch, F.~Bruckmann, and
  T.~Wettig,\href{http://dx.doi.org/10.1007/JHEP10(2013)140}{ {\em JHEP} {\bf
  1310} (2013) 140} [\href{http://arxiv.org/abs/1307.1416}{{\tt
  arXiv:1307.1416}}].

\bibitem{Aarts:2008rr}
G.~Aarts and I.-O.
  Stamatescu,\href{http://dx.doi.org/10.1088/1126-6708/2008/09/018}{ {\em JHEP}
  {\bf 0809} (2008) 018} [\href{http://arxiv.org/abs/0807.1597}{{\tt
  arXiv:0807.1597}}].

\bibitem{Aarts:2010gr}
G.~Aarts and K.~Splittorff,\href{http://dx.doi.org/10.1007/JHEP08(2010)017}{
  {\em JHEP} {\bf 1008} (2010) 017} [\href{http://arxiv.org/abs/1006.0332}{{\tt
  arXiv:1006.0332}}].

\bibitem{Nishimura:2015pba}
J.~Nishimura and
  S.~Shimasaki,\href{http://dx.doi.org/10.1103/PhysRevD.92.011501}{ {\em Phys.
  Rev.} {\bf D92} (2015) 011501} [\href{http://arxiv.org/abs/1504.08359}{{\tt
  arXiv:1504.08359}}].

\bibitem{Aarts:2013uxa}
G.~Aarts, L.~Bongiovanni, E.~Seiler, D.~Sexty, and I.-O.
  Stamatescu,\href{http://dx.doi.org/10.1140/epja/i2013-13089-4}{ {\em
  Eur.Phys.J.} {\bf A49} (2013) 89} [\href{http://arxiv.org/abs/1303.6425}{{\tt
  arXiv:1303.6425}}].

\bibitem{Bloch:2013qva}
J.~Bloch, F.~Bruckmann, and
  T.~Wettig,\href{http://pos.sissa.it/archive/conferences/187/194/LATTICE\%202013_194.pdf}{
  {\em PoS} (LATTICE 2013) 194} [\href{http://arxiv.org/abs/1310.6645}{{\tt
  arXiv:1310.6645}}].

\bibitem{Bloch:2015iha}
J.~Bloch and F.~Bruckmann, \href{http://arxiv.org/abs/1508.03522}{{\tt
  arXiv:1508.03522}}.

\bibitem{Mollgaard:2014mga}
A.~Mollgaard and
  K.~Splittorff,\href{http://dx.doi.org/10.1103/PhysRevD.91.036007}{ {\em Phys.
  Rev.} {\bf D91} (2015) 036007} [\href{http://arxiv.org/abs/1412.2729}{{\tt
  arXiv:1412.2729}}].

\bibitem{KeitaroNagata2015}
K.~Nagata, J.~Nishimura, and S.~Shimasaki, {\em PoS} (LATTICE 2015) 156.

\end{thebibliography}\endgroup
\bibliographystyle{jbJHEP_notitle}

\end{document}